\documentclass{article}
\usepackage{import}
\usepackage{preambles}
\usepackage{cite}
\numberwithin{equation}{section}
\allowdisplaybreaks

\begin{document}
\hfuzz=100pt
\title{String theory, $\mathcal{N}=4$ SYM and Riemann hypothesis}

\author{
\Large 
Masazumi Honda$^{1}$\footnote{masazumi.honda(at)yukawa.kyoto-u.ac.jp}, \quad
Takuya Yoda$^{2}$\footnote{t.yoda(at)gauge.scphys.kyoto-u.ac.jp}
\vspace{1em} \\
\\
{$^{1}$ \small{\it Center for Gravitational Physics, Yukawa Institute for Theoretical Physics}}\\ 
{\small{\it Kyoto University, Sakyo-ku, Kyoto 606-8502, Japan}} \\
{$^{2}$ \small{\it Department of Physics, Kyoto University, Sakyo-ku, Kyoto 606-8502, Japan}}
}

\date{\small{April 2022}}

\maketitle
\thispagestyle{empty}
\centerline{}

\begin{abstract}
We discuss new relations among string theory, 
four-dimensional $\mathcal{N}=4$ supersymmetric Yang-Mills theory (SYM)
and the Riemann hypothesis.
It is known that
the Riemann hypothesis is equivalent to
an inequality for the sum of divisors function $\sigma (n)$.
Based on previous results in literature,
we focus on 
the fact that $\sigma (n)$ appears 
in a problem of counting supersymmetric states 
in the $\mathcal{N}=4$ SYM with $SU(3)$ gauge group:
the Schur limit of the superconformal index
plays a role of a generating function of $\sigma (n)$.
Then assuming the Riemann hypothesis gives 
bounds on information on the $1/8$-BPS states in the $\mathcal{N}=4$ SYM.
The AdS/CFT correspondence further connects 
the Riemann hypothesis to the type IIB superstring theory on $AdS_5 \times S^5$.
In particular, the Riemann hypothesis implies
a miraculous cancellation 
among Kaluza-Klein modes of the supergravity multiplet and
D3-branes wrapping supersymmetric cycles in the string theory.
We also discuss possibilities
to gain new insights on the Riemann hypothesis from the physics side.

\end{abstract}

\vfill
\noindent

YITP-22-30, KUNS2920

\renewcommand{\thefootnote}{\arabic{footnote}}
\setcounter{footnote}{0}

\newpage
\pagenumbering{arabic}
\setcounter{page}{1}

\clearpage
\section{Introduction}
The Riemann hypothesis is the conjecture that
the Riemann zeta function $\zeta (s)$ has nontrivial zeroes
only along the line ${\rm Re}(s)=1/2$,
where the trivial zeroes are at $s=-2n$ with $n\in\mathbb{Z}_+$.
This is an important problem especially in number theory
as the zeroes of the zeta function have significant information on prime number distribution,
and indeed one of the millennium prize problems by Clay Mathematics Institute.
While this is originally a purely mathematical problem,
its physical realization 
-- since the Hilbert-P\'{o}lya conjecture \cite{odlyzko,montgomery} --
has been discussed in various problems of physics 
(see e.g.~review \cite{Schumayer:2011yp,Wolf:2014ulr,numtheroyphys}),
including quantum mechanics, statistical mechanics,
random matrix theory 
and string theory \cite{He:2010jh,He:2015jla,Cacciatori:2010js,Angelantonj:2010ic,Cacciatori:2011fc}.
In this paper,
we point out new relations among the Riemann hypothesis,
string theory and gauge theory.

\begin{figure}[t]
\centering
\includegraphics[scale=0.45]{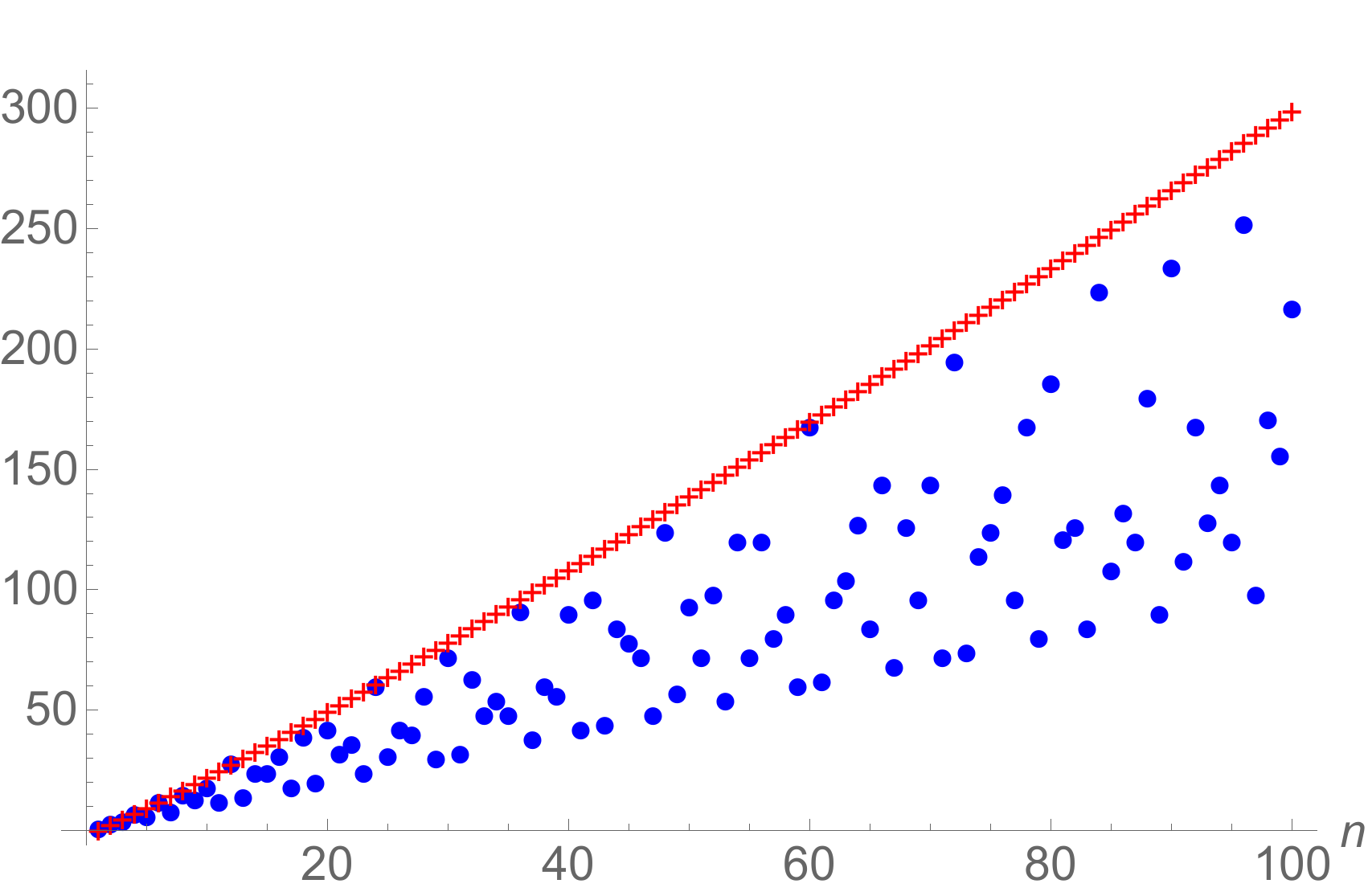}
\caption{%
A comparison of $\sigma (n)$ (blue circle) and $H_n +e^{H_n} \log{H_n}$ (red plus) in \eqref{eq:ineq} up to $n=100$.
}
\label{fig:inequality}
\end{figure}

There are various equivalent statements of the Riemann hypothesis.
Here we focus on one of them \cite{Lagarias}
in terms of the sum of divisors function $\sigma (n)$ :
\begin{\eq}
\sigma (n) :=\sum_{d|n} d ,
\end{\eq}
where $\sum_{d|n}$ stands for the sum over the integers $d$ which divide $n$.
It was proven in \cite{Lagarias} that 
the Riemann hypothesis is equivalent to 
the following inequality\footnote{
This inequality was derived from the Robin's inequality \cite{Robin}:
$\sigma (n) \leq e^\gamma n \log{\log{n}}$ for $n \geq 5041$
with the Euler constant $\gamma$.
} (c.f.~fig.~\ref{fig:inequality}) \cite{Lagarias}:
\begin{\eq}
\sigma (n) 
\leq H_n +e^{H_n} \log{H_n} \quad {\rm for}\ ^\forall n\in \mathbb{Z}_+ ,
\label{eq:ineq}
\end{\eq}
where the equality holds only for $n=1$
and 
$H_n$ is the harmonic number defined by
\begin{\eq}
H_n := \sum_{k=1}^n \frac{1}{k} .
\end{\eq}

In this paper we discuss
implications of the inequality \eqref{eq:ineq} to physics:
gauge theory and string theory.
We start with pointing out that
there is a physical quantity in the four-dimensional $\mathcal{N}=4$ supersymmetric Yang-Mills theory (SYM)
which takes a form of a generating function of the sum of divisors function $\sigma (n)$
according to previous results in the literature \cite{Bourdier:2015wda,Bourdier:2015sga,Kang:2021lic,Beem:2021zvt}.
The quantity is a supersymmetric (SUSY) index
called the Schur index in the $\mathcal{N}=4$ SYM with gauge group $SU(3)$.
The Schur index is a special limit of a four-dimensional superconformal index 
and counts the difference between the numbers of bosonic and fermionic SUSY states 
annnihilated by two Poincare supercharges: 
\begin{\eq}
I_{\rm Schur} (q)
= {\rm Tr}_{\rm SUSY}  \Bigl[  (-1)^F  q^{2(E-R)} \Bigr] ,
\end{\eq}
where $F$ is fermion number, $E$ is energy and $R$ is an $R$-charge.
The Schur index of the $SU(3)$ $\mathcal{N}=4$ SYM is known to take
the following form \cite{Bourdier:2015wda,Bourdier:2015sga,Kang:2021lic,Beem:2021zvt,Pan:2021mrw}
\begin{\eq}
I_{SU(3)} (q) =  \sum_{n=1}^\infty \sigma (n) q^{2(n-1)} ,
\label{eq:SU3}
\end{\eq}
which is the generating function of $\sigma (n)$.
Therefore $\sigma (n)$ corresponds to the difference between the numbers of
bosonic and fermionic $1/8$-BPS states with the quantum number $E-R =n-1$.
This enables us to rephrase the Riemann hypothesis 
in the language of the $\mathcal{N}=4$ SYM.
In particular 
the Riemann hypothesis claims
bounds on information on the $1/8$-BPS states in the $SU(3)$ $\mathcal{N}=4$ SYM.
Furthermore
the AdS/CFT correspondence \cite{Maldacena:1997re} gives an interpretation of the Schur index 
from the dual string theory \cite{Arai:2020qaj} and 
then leads us to relations between the Riemann hypothesis and string theory.
We argue that
the Riemann hypothesis implies
a huge cancellation 
among Kaluza-Klein modes of the supergravity multiplet and
D3-branes wrapping supersymmetric cycles in the type IIB superstring theory on $AdS_5 \times S^5$.
We also discuss possibilities
to gain some insights on the Riemann hypothesis from the physics side.

This paper is organized as follows.
In sec.~\ref{sec:Schur}, we discuss the Schur index of the $\mathcal{N}=4$ SYM
and its relation to the Riemann hypothesis.
In sec.~\ref{sec:string}, we discuss its interpretations from the string theory 
via the AdS/CFT correspondence.
Sec.~\ref{sec:discussion} is devoted to discussions.

\section{$\mathcal{N}=4$ SYM and Riemann hypothesis}
\label{sec:Schur}
In this paper we study a SUSY index called the Schur index 
which is a limit of a four-dimensional superconformal index.
In general, 
SUSY indices can be defined through partition functions of SUSY theories on manifolds including $S^1$.
The most elementary one is the Witten index \cite{Witten:1982df} of SUSY quantum mechanics given by
\begin{align}
I_W   &= \Tr\left[ (-1)^F e^{-\beta H} \right] ,
\end{align}
where the trace is over the entire Hilbert space,
$\beta$ is a radius of $S^1$
and $F$ is the fermion number operator.
While it apparently depends on $\beta$,
it is known that
there are typically the same numbers of bosonic and fermionic non-SUSY states,
which always have non-zero energies,
and only SUSY states with zero energy give net contributions to the Witten index:
\begin{align}
I_W   &= \Tr_{\rm SUSY}\left[ (-1)^F \right] ,
\end{align}
where the trace is now over the SUSY states.
Therefore the Witten index is independent of $\beta$
and gives a difference between the numbers of bosonic and fermionic SUSY states.
More generally 
one can consider a refinement by chemical potentials of global symmetries 
which can be both internal symmetries and isometries of space,
and generic SUSY index is schematically represented as
\begin{align}
    I_{\rm SUSY} \left( \{ \mu_i \} \right)
    &= \Tr_{\rm SUSY}\left[ (-1)^F \prod_{i} e^{\mu_i Q_i } \right] ,
\end{align}
where $Q_i$'s are the conserved charges of global symmetries commuting 
with a part of supercharges.
Expanding this leads us to
\begin{align}
I_{\rm SUSY} \left( \{ \mu_i \} \right)
    &= \sum_{\{Q_i\}} \left( d_{\text{B}}(\{Q_i\})-d_{\text{F}}(\{Q_i\}) \right) \prod_{i}e^{\mu_i Q_i} ,
\end{align}
where $d_{\text{B/F}}(\{Q_i\})$ denotes the number of bosonic/fermionic SUSY states
with the quantum numbers $\{Q_i\}$.
One advantage to study SUSY index is that
it is SUSY invariant and can be often computed exactly.
Therefore SUSY index has been used to explore strong coupling behavior of SUSY theories.

Superconformal index is a Witten index \cite{Witten:1982df} 
of a radially quantized SUSY theory (i.e.~on $\mathbb{R}\times S^{d-1}$)
refined by chemical potentials of global symmetries.
It is also related to counting of SUSY operators via the state/operator correspondence in conformal field theory as $\mathbb{R}\times S^{d-1}$ is conformally flat\footnote{
Strictly speaking, superconformal index can be defined even for non-conformal theory and the state/operator correspondence holds only for the superconformal cases.
}.
In the case of 4d $\mathcal{N}=2$ SUSY theory,
the superconformal index is defined as\footnote{
This representation is associated with certain supercharge 
in the notation of \cite{Gadde:2011uv} (up to rescaling of the fugacities). 
In 4d $\mathcal{N}=1$ language, it is written as
\[
I_{\rm SCI}
= {\rm Tr}_{\rm SUSY}\Bigl[  (-1)^F 
 p^{j_1 +j_2+\frac{1}{2}r_{\mathcal{N}=1}} q^{-j_1 +j_2 +\frac{1}{2}r_{\mathcal{N}=1}} v^{\frac{3}{2}r_{\mathcal{N}=1}-3R} 
\Bigr] ,
\]
where $v=(pq)^{2/3} t^{-1}$.
}
\begin{\eq}
I_{\rm SCI} 
= {\rm Tr}_{\rm SUSY} \Bigl[  (-1)^F  p^{j_1 +j_2-r} q^{2(-j_1 +j_2 -r)} t^{2(R+r)} \Bigr] ,
\end{\eq}
where $(j_1 , j_2 )$ is the Cartan part of the $S^3$ isometry $SO(4)\simeq SU(2)\times SU(2)$,
$R$ is the Cartan part of $SU(2)_R$ and $r$ is the $U(1)_r$ charge.
The trace is over the SUSY states annihilated by one Poincare supercharge with
\begin{\eq}
 E -2j_2 -2R +r =0 .
\label{eq:bound_index1}
\end{\eq}
Then the Schur index is defined as the following limit of the superconformal index\footnote{
More precisely, minimally we need $q=t$. 
Although $p$ can be arbitrary for this case, the index is known to be independent of $p$.
}
\begin{\eq}
p\rightarrow 1 ,\quad q=t ={\rm fixed} ,
\end{\eq}
which amounts to counting the SUSY states annihilated by two Poincare supercharges 
with
\begin{\eq}
E +j_1 -j_2 -2R=0 ,
\end{\eq}
in addition to \eqref{eq:bound_index1}.
The trace representation of the Schur index is\footnote{
It has been conjectured \cite{Cordova:2015nma} that
the Schur index can be also computed from BPS spectrum on Coulomb branch
captured by the wall-crossing formula in \cite{Kontsevich:2008fj}. 
}
\begin{\eq}
I_{\rm Schur} (q)
= {\rm Tr}_{\rm SUSY} \Bigl[  (-1)^F  q^{2(-j_1 +j_2 +R)} \Bigr] 
= {\rm Tr}_{\rm SUSY}  \Bigl[  (-1)^F  q^{2(E-R)} \Bigr] .
\end{\eq}

From now on,
we focus on the four-dimensional $\mathcal{N}=4$ supersymmetric Yang-Mills theory 
which is the low energy worldvolume theory of D3-branes in the type IIB string theory.
The $\mathcal{N}=4$ SYM consists of an $\mathcal{N}=2$ vector multiplet and one adjoint hypermultiplet in 4d $\mathcal{N}=2$ language.
It is known that
the superconformal index of any $\mathcal{N}=1$ Lagrangian theory
has a finite dimensional integral representation with dimensions ${\rm rank}(G)$
\cite{Kinney:2005ej,Romelsberger:2005eg,Dolan:2008qi,Nawata:2011un,Imamura:2011uw,Assel:2014paa},
which is physically an integration over gauge holonomy along $S^1$.
For the case of the Schur index in the $\mathcal{N}=4$ SYM with gauge group $G$,
it is given by
\begin{\eq}
I_{G} (q)
= \frac{1}{|W(G)|} 
\left(  \prod_{n=0}^\infty \frac{1 -q^{2n+2 }  }{1 -q^{2n+1 }  } \right)^{2{\rm rank}(G)}
\left( \prod_{j=1}^{{\rm rank}(G)} \oint \frac{dz_j}{2\pi i z_j} \right) \ 
 \prod_{\alpha \in {\rm root}_{\neq 0} }  
 \frac{\theta (z^\alpha  ; q^2) }{ \theta (qz^\alpha  ;q^2 )} ,
\end{\eq}
where $|W(G)|$ denotes the rank of the Weyl group of $G$ and
\begin{\eq}
\theta (z ;q^2 ) 
:= \prod_{n=0}^\infty \left( 1-zq^{2n} \right) \left( 1-z^{-1} q^{2(n+1)} \right) .
\end{\eq}

There are various works 
which wrote down
a closed form of the Schur index of the $\mathcal{N}=4$ SYM 
for various gauge groups \cite{Bourdier:2015wda,Bourdier:2015sga,Kang:2021lic,Beem:2021zvt,Pan:2021mrw}.
Here we are particularly interested in the $G=SU(3)$ case:
\begin{\eq}
I_{SU(3)} (q)
=\frac{q^{-2}}{24} \left( 1- E_2 ( q ) \right) ,
\label{eq:Eisen}
\end{\eq}
where $E_2 (q ) $ is the Eisenstein series $E_{2k} (q )$ with $k=1$ defined by
\begin{\eq}
E_{2k} (q )
:= \frac{1}{2\zeta (2k)}\sum_{(m,n) \neq (0,0)} \frac{1}{(m +n\tau )^{2k}} 
\quad {\rm with}\ \ q^2 := e^{2\pi i\tau }  .
\end{\eq}
The Eisenstein series is also known to have the following representation
\begin{\eq}
E_{2} (q )
= 1 -24\sum_{n=1}^\infty \frac{nq^{2n}}{1- q^{2n}}
= 1 -24\sum_{n=1}^\infty \sigma (n) q^{2n} .
\end{\eq}
Plugging this into \eqref{eq:Eisen}, 
we find
\[
I_{SU(3)} (q) =  \sum_{n=1}^\infty \sigma (n) q^{2(n-1)} .
\]
Thus the Schur index of the $SU(3)$ $\mathcal{N}=4$ SYM is
the generating function of the sum of divisors function $\sigma (n)$.
The sum of divisors function $\sigma(n)$
directly corresponds to the difference between the numbers of
bosonic and fermionic $1/8$-BPS states with the quantum number $E-R =n-1$.
Therefore
the Riemann hypothesis \eqref{eq:ineq} is physically rephrased as follows:
the difference of the numbers of 
the bosonic and fermionic $1/8$-BPS states with $E-R =n-1$ 
in the $SU(3)$ $\mathcal{N}=4$ SYM
is less than or equal to $ H_n +e^{H_n} \log{H_n} $.

\section{String theory and Riemann hypothesis}
\label{sec:string}
The AdS/CFT correspondence tells us that
the $SU(N)$ $\mathcal{N}=4$ SYM is dual to the type IIB superstring theory on $AdS_5 \times S^5$ \cite{Maldacena:1997re}.
A natural question is
whether the relation between the $\mathcal{N}=4$ SYM and the Riemann hypothesis 
can be translated to the string theory language.
Here we provide such a relation
based on a recent proposal on an interpretation of 
the Schur index of the $\mathcal{N}=4$ SYM from the gravity side for finite $N$ \cite{Arai:2020qaj} (see also \cite{Aharony:2021zkr,Gaiotto:2021xce,Imamura:2021ytr}).

Before that, let us recall the dictionary of the AdS/CFT correspondence \cite{Maldacena:1997re}.
The plank length $\ell_p$ and the string length $\ell_s$
in the unit of the AdS radius $R$ are 
related to $N$ and the 't Hooft coupling $\lambda$ by
\begin{\eq}
 \frac{\ell_p}{R} = (4\pi N)^{1/4} ,\quad \frac{\ell_s}{R} = \lambda^{-1/4} .
\end{\eq}
In the large $N$-limit and strong 't Hooft coupling limit,
the $\mathcal{N}=4$ SYM should be approximated by the type IIB supergravity on $AdS_5 \times S^5$.
Correspondingly, it is known that
the superconformal index of the $\mathcal{N}=4$ SYM in the large-$N$ limit\footnote{
The 't Hooft coupling is irrelevant in this discussion
since the superconformal index is insensitive to $\lambda$.
} 
is described by Kaluza-Klein (KK) modes of the supergravity multiplet \cite{Kinney:2005ej}.
When $N$ is finite, 
the dictionary tells us that quantum gravity effect becomes important.
In particular, rewriting the dictionary in terms of the D3-brane tension $T_{\rm D3}$ as
\begin{\eq}
\frac{T_{\rm D3}}{R^4} = \frac{N}{2\pi^2} ,
\end{\eq}
it is clear that the D3-branes have finite tensions for finite $N$ and
physical quantities likely get effects of extended D3-branes.

The proposal in \cite{Arai:2020qaj} manifests the above intuitions for the Schur index.
The authors \cite{Arai:2020qaj} first decomposed 
the Schur index of the $U(N)$ $\mathcal{N}=4$ SYM as
\begin{\eq}
I_{U(N)} (q) 
= I_{\rm KK } (q) \sum_{n=0}^\infty I_n^{BDF} (q;N) ,
\end{\eq}
where $I_{\rm KK } (q)$ is the Schur index for the $U(\infty )$ case,
equivalent to the contribution from the KK modes\footnote{
In terms of plethystic exponential, it is written as
$I_{\rm KK } (q)$ $=$ ${\rm Pexp} \left( \frac{2q}{1-q} -\frac{q^2}{1-q^2} \right) $.
}:
\begin{\eq}
I_{\rm KK } (q)
= \prod_{n=1}^\infty \frac{1 }{(1-q^n ) (1-q^{2n-1} )} .
\end{\eq}
The second factor $I_n^{BDF} (q;N)$, based on the exact computation in \cite{Bourdier:2015sga},
is given by
\begin{\eq}
I_n^{BDF} (q;N)
= \sum_{n=0}^\infty  (-1)^n \left( _{N+n}C_N +\ _{N+n-1}C_N \right) q^{nN+n^2}.
\end{\eq}
The proposal in \cite{Arai:2020qaj} is 
about an interpretation of the object $I_n^{BDF} (q;N)$ from the string theory side:
it comes from contributions from D3-branes wrapping SUSY cycles.
Concretely,
there are $1/8$-BPS brane configurations on the string theory side \cite{Biswas:2006tj}
which are given by the intersection of a holomorphic surface $h(X,Y,Z) =0$
and $S^5 =\{ (X,Y,Z) | |X|^2 +|Y|^2 +|Z|^2 =1\}$ \cite{Mikhailov:2000ya}.
Then the authors in \cite{Arai:2020qaj} proposed 
\begin{\eq}
I_n^{BDF} (q;N)
= \sum_{k=0}^n I_{(n-k ,k)} (q ;N) ,
\end{\eq}
where $I_{(n_1,n_2 )}$ denotes the contribution from $n_1$ D3-branes wrapped on $X=0$
and $n_2$ D3-branes wrapped on $Y=0$ (see \cite{Arai:2020qaj} for details).
To get the result for the $SU(N)$ case,
we can simply use the following relation
\begin{\eq}
I_{SU(N)} (q)
= I_{U(N)} (q) \prod_{n=0}^\infty \frac{(1 -q^{2n+1 }  )^2 }{(1 -q^{2n+2 } )^2  } . 
\end{\eq}
Then we find the Schur index for the $SU(N)$ case as
\begin{\eq}
I_{SU(N)} (q) 
= \Biggl[ \prod_{n=1}^\infty \frac{1 }{(1 -q^{2n } )^3  } \Biggr] 
\sum_{n=0}^\infty I_n^{BDF} (q;N) .
\label{eq:SUN}
\end{\eq}
Note that 
the first factor is the same as the generating function for the number of $M$-colored partitions\footnote{
It may be interesting to note that
this factor is formally the same as the Nekrasov instanton partition function \cite{Nekrasov:2002qd}
of $U(M)$ $\mathcal{N}=2^\ast$ theory 
with vanishing equivariant mass \cite{Pestun:2007rz,Okuda:2010ke}
although the parameter $q$ here is not apparently related to complex gauge coupling.
} 
with $M=3$:
\begin{\eq}
\prod_{n=1}^\infty \frac{1 }{(1 -z^n )^M  }
= \sum_{\vec{Y}} z^{|\vec{Y}|} ,
\end{\eq}
where $\vec{Y}$ denotes the set of $M$ Young diagrams of $U(M)$ and
$|\vec{Y}|$ is the total number of boxes of $\vec{Y}$.

To see a relation to the Riemann hypothesis, let us go back to the $SU(3)$ case:
\begin{\eqa}
I_{SU(3)} (q) 
&=& \Biggl[ \prod_{n=1}^\infty \frac{1 }{(1 -q^{2n } )^3  } \Biggr] 
\sum_{n=0}^\infty I_n^{BDF} (q;3)  \NN\\
&=& \Biggl[ \prod_{n=1}^\infty \frac{1 }{(1 -q^{2n } )^3  } \Biggr] 
\sum_{n=0}^\infty  (-1)^n \frac{(n+1)(n+2)(2n+3)}{6} q^{3n+n^2} .
\label{eq:SU3_2}
\end{\eqa}
Compared with the expression \eqref{eq:SU3},
the RHS in \eqref{eq:SU3_2} must generate the sum of divisors function $\sigma (n)$
and the Riemann hypothesis claims that
it is bounded by $H_n +e^{H_n} \log{H_n}$, 
which is asymptotically $\mathcal{O}(n\log{\log{n}})$ for large $n$.
This implies that
we have a kind of miraculous cancellations between the first and second factors in \eqref{eq:SU3_2}:
the first factor generates factorially growing coefficients with positive definite sign
while the second one gives $\mathcal{O}(n^3 )$ coefficients with alternate signs.
Let us explicitly demonstrate it.
The first factor in \eqref{eq:SU3_2} is expanded as
\begin{\eqa}
\prod_{n=1}^\infty \frac{1 }{(1 -q^{2n } )^3}
&=& 1 +3 q^2 +9 q^4+22 q^6+51 q^8+108 q^{10}+221 q^{12}+429 q^{14}
 +810 q^{16}+1479 q^{18} \NN\\
&& +2640 q^{20}+4599 q^{22}+7868 q^{24}+13209 q^{26}+21843 q^{28} +35581q^{30} 
   +\mathcal{O} \left(q^{31}\right) ,
\end{\eqa}
while the second factor is expanded as
\begin{\eq}
\sum_{n=0}^\infty I_n^{BDF} (q;3)
= 1-5 q^4+14 q^{10}-30 q^{18}+55 q^{28}  +\mathcal{O} \left(q^{40}\right) .
\end{\eq}
Multiplying the two factors, we find
\begin{\eqa}
I_{SU(3)} (q)
&=& 1+3 q^2+4 q^4+7 q^6+6 q^8+12 q^{10}+8 q^{12}+15 q^{14}+13 q^{16}+18 q^{18} \NN\\
&& +12 q^{20} +28   q^{22}+14 q^{24}+24 q^{26}+24 q^{28}+31 q^{30}
+\mathcal{O} \left( q^{31} \right) ,
\end{\eqa}
whose coefficients are indeed the sum of divisors function $\sigma (n)$.
The cancellations may imply that
there is a more appropriate string language which describes the net contributions to the Schur index more directly.
This might be something like bound states of the KK modes and the wrapping D3-branes.

\section{Discussions}
\label{sec:discussion}
In this paper 
we have pointed out the new relations among the $\mathcal{N}=4$ SYM, string theory and
Riemann hypothesis.
We started with noting the fact that
the Schur index of the $SU(3)$ $\mathcal{N}=4$ SYM is 
the generating function of the sum of divisors function $\sigma (n)$ \cite{Bourdier:2015wda,Bourdier:2015sga,Kang:2021lic,Beem:2021zvt,Pan:2021mrw}, 
which appears in the statement \eqref{eq:ineq} equivalent to the Riemann hypothesis.
In this identification, 
$\sigma (n)$ corresponds to the difference between the numbers of
bosonic and fermionic $1/8$-BPS states with the quantum number $E-R =n-1$,
and the Riemann hypothesis claims the upper bound on that.
Then the AdS/CFT correspondence allows us to interpret it from the viewpoint of the string theory.
According to the proposal in \cite{Arai:2020qaj},
the Schur index has contributions from the KK modes of the supergravity multiplet and wrapping D3-branes,
and their complicated combinations give the sum of divisors function after the miraculous cancellations.

An immediate question is whether it is just a coincidence or
there is a deep physical origin.
One might also ask why the $SU(3)$ $\mathcal{N}=4$ SYM is ``selected''.
At least the structure discussed in this paper does not seem unique for the $SU(3)$ $\mathcal{N}=4$ SYM 
because there is another theory called $\hat{D}_4 (SU(3))$ $\mathcal{N}=2$ superconformal theory 
which has the same Schur index as the $SU(3)$ $\mathcal{N}=4$ SYM after the rescaling $q\rightarrow q^2$ \cite{Kang:2021lic}.
Regardless of these questions,
it would be extremely interesting
if one can give some implications on the Riemann hypothesis from the physics side.
One possible scenario would be that 
some constraints by physical requirements could (un)support the Riemann hypothesis.
For example, 
while unitarity bound would be too obvious,
we could consider constraints by quasi-modularity \cite{Razamat:2012uv,Beem:2021zvt},
information theoretic inequalities \cite{Zhou:2016kcz,Zhou:2015cpa}
and/or something else for the Schur index.

It may be also useful to look at the Schur index from different viewpoints.
In this paper we have discussed the Schur index from the viewpoint of SUSY states counting. 
It is also known that
the superconformal index is proportional 
to a SUSY partition function on a space with a topology of $S^1 \times S^3$:
\begin{\eq}
Z_{S^1 \times S^3} = e^{-E_{\rm SUSY}} I_{\rm SCI} ,
\end{\eq}
where $E_{\rm SUSY}$ is so-called supersymmetric Casimir energy \cite{Assel:2015nca}. 
Then the AdS/CFT correspondence tells us that
this is dual to the partition function of the type IIB superstring theory on $AdS_5 \times S^5$.
It might be helpful to interpret the relation to the Riemann hypothesis from this viewpoint.

The Schur index of the $\mathcal{N}=4$ SYM is also known to correspond 
to physical quantities in other theories.
First,
it is known that 
the Schur index of the $SU(N)$ $\mathcal{N}=4$ SYM corresponds 
to the partition function of the 2d $SU(N)$ $q$-deformed Yang-Mills theory on torus in the zero area limit \cite{Gadde:2011ik,Gadde:2011uv} 
in a similar spirit to the AGT relation \cite{Alday:2009aq}.
Because the $q$-deformed YM appears also in a problem of counting BPS black holes 
in type II superstring on a local Calabi-Yau threefold \cite{Aganagic:2004js},
this gives another connection between the string theory and Riemann hypothesis.
Second, 
it is also expected that
the Schur index of the $\mathcal{N}=4$ SYM with gauge group $G$ corresponds 
to a torus partition function of a chiral algebra called an $\mathcal{N}=4$ super-$\mathcal{W}$ algebra with ${\rm rank}(G)$ generators \cite{Beem:2013sza}.
Because there is also an argument which derives a chiral algebra structure 
from the SUSY localization on $AdS_5$ \cite{Bonetti:2016nma},
it may also imply a new relation between the string theory and Riemann hypothesis.
These viewpoints may give some insights on the Riemann hypothesis.

There may be also a connection to black hole microstate counting
as the superconformal index is known to account for SUSY black hole entropy on the string theory side \cite{Choi:2018hmj,Benini:2018ywd,Honda:2019cio,ArabiArdehali:2019tdm} 
similar to the spirit of the seminal work by Strominger-Vafa \cite{Strominger:1996sh}.
Although the Schur index itself does not seem to capture the black hole entropy due to cancellations,
the Macdonald index, 
which counts the same sector as the Schur index but with a refinement,
has a limit to reproduce the black hole entropy \cite{Choi:2018hmj}.
To explore such a connection,
it may be useful to consider the ``Cardy limit'' \cite{DiPietro:2014bca,Ardehali:2015bla,DiPietro:2016ond}.
A connection between the black hole and the Riemann hypothesis might be expected 
also by noting the two intuitions that 
black holes are related to chaos (see e.g.~\cite{Sekino:2008he,Shenker:2013pqa}) 
and chaos is related to the Riemann hypothesis (see e.g.~\cite{Schumayer:2011yp,Wolf:2014ulr, numtheroyphys}).

In this paper 
we have focused on connections between string theory and
the Riemann hypothesis for the Riemann zeta function.
It would be illuminating to explore connections for other types of zeta functions.
For example,
while we have focused on the $SU(3)$ case of the $\mathcal{N}=4$ SYM,
the Schur indices for other gauge groups may be related to more general zeta functions.
One possible hint may be the fact that
there appear Macmahon's generalized sum of divisors function \cite{MacMahon}
and more general Eisenstein series for $SU(N)$ case \cite{Kang:2021lic,Beem:2021zvt}.

\subsection*{Acknowledgment}
This work was inspired by an April Fool's joke tweet made by T.~Y.  in 2021 while this work itself is not a joke. 
M.~H. would like to than So Matsuura for some conversations.
M.~H. is supported by MEXT Q-LEAP, JST PRESTO Grant Number JPMJPR2117
and JSPS Grant-in-Aid for Transformative Research Areas (A) JP21H05190.
T.~Y. is supported by Sasakawa Scientific Research Grant from The Japan Science Society,
and by JST SPRING Grant Number JPMJSP2110.

\bibliographystyle{utphys}
\bibliography{RH}

\end{document}